\documentclass[twocolumn,prc,superscriptaddress]{revtex4-2}
\usepackage{graphicx,color}
\usepackage{float}
\usepackage{bm}
\usepackage[pdftex,colorlinks=true, linkcolor = blue, citecolor=blue,urlcolor=blue, bookmarksnumbered=true, bookmarksopen=true]{hyperref}

\begin{document}

\title{Updating GEANIE $^{239}$Pu prompt $\gamma$-ray experimental data using modern Hauser-Feshbach fission fragment decay model}

\author{T. Kawano}
\email{kawano@lanl.gov}
\affiliation{Los Alamos National Laboratory, Los Alamos, NM 87545, USA}

\author{A. E. Lovell}
\affiliation{Los Alamos National Laboratory, Los Alamos, NM 87545, USA}

\author{P. Talou}
\affiliation{Los Alamos National Laboratory, Los Alamos, NM 87545, USA}
\affiliation{Stardust Science Labs, Santa Fe, New Mexico 87507, USA}

\author{L. A. Bernstein}
\affiliation{Department of Nuclear Engineering, University of California, Berkeley, CA 94720, USA}
\affiliation{Lawrence Berkeley National Laboratory, Berkeley, CA 94720, USA}

\date{\today}

\begin{abstract}

We calculate fission $\gamma$ rays for neutron-induced reactions on
$^{239}$Pu with the Hauser-Feshbach fission fragment decay model. By
applying the calculated fission $\gamma$ rays as a background
contribution, the historical $^{239}$Pu(n,$x$n$\gamma$) reaction cross
section data measured by the GEANIE (GErmanium Array for Neutron
Induced Excitations) spectrometer are corrected. The correction also
includes other (n,$x$n) reactions that have very similar energies to
the $\gamma$ lines reported by GEANIE. In many cases, the original
GEANIE data are strongly reduced and they become much closer to the
statistical Hauser-Feshbach model predictions. The total inelastic
scattering, (n,2n), and (n,3n) cross sections are inferred based on
the corrected GEANIE data, and compared with available experimental
data as well as the statistical model calculations. Expected
$\gamma$-ray energy spectra for neutron-induced measurements on
$^{239}$Pu are also discussed.
\end{abstract}
\maketitle

\section{Introduction}
\label{sec:introduction}

GEANIE (GErmanium Array for Neutron Induced Excitations) is a
high-resolution $\gamma$-ray spectrometer once installed at LANSCE
(Los Alamos Neutron Science Center) at Los Alamos National Laboratory.
In 2001, Younes {\it et al.}~\cite{Younes2001} reported fission
$\gamma$-ray measurement for the neutron-induced reaction on $^{235}$U
with GEANIE. In 2002, Bernstein {\it et al.}~\cite{UCRL-ID-140308}
reported the $\gamma$-ray production cross section for the
$^{239}$Pu(n,$x$n) reactions, and inferred the $^{239}$Pu(n,2n) cross
section~\cite{Bernstein2002} by employing the statistical
Hauser-Feshbach model with the GNASH code~\cite{LA12343,
  LA-UR-99-2885}. This experiment and data analysis were particularly
challenging due to the high $\gamma$-ray background produced by
fission and difficulties in incorporating the fission channel into the
Hauser-Feshbach calculation at that time.

On the experimental side after a quarter of a century, similar
experimental setups have benefitted the partial $\gamma$-ray
measurement technique, {\it e.g.,} the GENESIS (Gamma Energy Neutron
Energy Spectrometer for Inelastic Scattering)
project~\cite{Gordon2024} at LBNL (Lawrence Berkeley National
Laboratory) and the GRAPhEME (GeRmanium array for Actinides PrEcise
MEasurements) setup~\cite{Kerveno2021} in GELINA (Geel Electron
LIN-ear Accelerator) at JRC (Joint Research Centre) in Belgium. On the
theory side, the theoretical modeling for nuclear reactions on
actinides has improved significantly in the last decade. There are
several important achievements in this field: a rigorous treatment of
nuclear deformation by the coupled-channels Hauser-Feshbach
formalism~\cite{Kawano2016,Kawano2021}, a quantum mechanical effect on
the $\gamma$-ray production~\cite{Dashdorj2007,Kerveno2021}, and a new
barrier penetrability calculation in the fission
channel~\cite{Kawano2024}.

The (n,$x$n$\gamma$) measurements for actinides might include some
$\gamma$ rays originated from fission fragments. In the past it was
impossible to estimate the prompt fission $\gamma$ rays, because the
fission fragment yields at various incident neutron energies were
unknown. The fission $\gamma$-ray data were represented by a lumped
energy spectrum, in which information on individual $\gamma$ lines is
unavailable. Once nuclear fission takes place, two highly excited
fission fragments decay by emitting prompt neutrons and $\gamma$ rays.
Since a large number of $\gamma$ rays are produced by all the fission
fragments, these $\gamma$ rays could have similar energy to those
produced by the (n,$x$n$\gamma$) reactions that we want to
measure. Okumura {\it et al.}~\cite{Okumura2018,Okumura2022} developed
the Hauser-Feshbach Fission Fragment Decay (HF$^3$D) model to estimate
the energy dependence of fission observables, and Lovell {\it et
  al.}~\cite{Lovell2021} implemented the multi-chance fission process
in the HF$^3$D model. These developments allow us to calculate
emission probabilities of thousands of prompt $\gamma$ rays as a
function of incident neutron energy.

This paper aims at revisiting the historical GEANIE experiment by
introducing recent developments of theoretical nuclear physics, and
exploring possible corrections to the reported GEANIE data for
$^{239}$Pu. In fact, it is known that the GNASH deduced
$^{239}$Pu(n,2n) cross sections from partial $\gamma$-ray production
cross section~\cite{Bernstein2002} are systematically lower than other
experimental data measured through the activation technique, such as
Lougheed {\it et al.}~\cite{Lougheed2002}. GEANIE also measured the
(n,n'$\gamma$) and (n,3n$\gamma$), although these data have never been
employed to estimate the (n,n') and (n,3n) cross sections so far.
Inferring (n,$x$n) from the partial $\gamma$-ray production data
relies on how the background component produced by other reactions is
characterized. The prompt $\gamma$-ray measurements for actinides are
particularly challenging because fission produces enormous $\gamma$
rays.  Discrimination of the reaction and fission $\gamma$ rays is
crucial to avoid unphysical corrections. Our method to eliminate the
fission $\gamma$ rays will be helpful for future designs of
experiments.

\section{Calculation Method}
\label{sec:calculation}

\subsection{Statistical model calculation}

We calculate neutron-induced reactions on $^{239}$Pu with the
statistical Hauser-Feshbach code CoH$_3$~\cite{Kawano2019}, and
collect all the $\gamma$ rays produced by the (n,n'), (n,2n), (n,3n)
and (n,4n) reactions. The coupled-channels optical potential of
Soukhovitskii {\it et al.}~\cite{Soukhovitskii2005} is employed for
the neutron transmission coefficients, and the so-called
Engelbrecht-Weidenm\"{u}ller transformation~\cite{Engelbrecht1973,
  Kawano2016} is performed to correctly take the width fluctuation
into account for the deformed nucleus.  The level density parameter of
$^{240}$Pu was slightly adjusted to reproduce the average $s$-wave
resonance spacing of 2.07~eV~\cite{Mughabghab2006}. The M1 scissors
mode~\cite{Mumpower2017} of the giant dipole resonance is included to
reproduce the average $\gamma$-ray width of
43~meV~\cite{Mughabghab2006}. The fission barrier parameters in the
CoH$_3$ fission model~\cite{Kawano2024} were adjusted to reproduce the
evaluated fission cross sections in ENDF/B-VIII.0~\cite{ENDF8} and
JENDL-5~\cite{JENDL5}, which represent all the available experimental
data.

\begin{figure}
 \resizebox{\columnwidth}{!}{\includegraphics{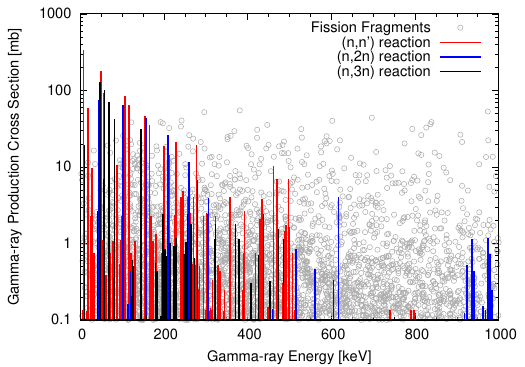}}
 \caption{Calculated $\gamma$-ray production cross sections for the
   neutron-induced reaction on $^{239}$Pu at 19.3~MeV as function of
   the $\gamma$ energy. The symbols
   represent $\gamma$ lines from the fission fragments. The vertical lines are from
   the (n,$x$n) reactions, where $x=1$, 2, and 3.}
 \label{fig:discgamma}
\end{figure}

For the $\gamma$ rays produced by the decay of prompt fission
fragments, we use the BeoH code that includes the multi-chance fission
process~\cite{Lovell2021}. BeoH is also based on the Hauser-Feshbach
theory, calculating decay of a compound nucleus from various excited
states for all fragments produced by fission. Because $\gamma$ rays
are emitted at any stage of the compound nucleus decay process, the
precursor of the $\gamma$ decay is not exactly the fission fragments
just after scission. However, in order to distinguish these compound
nuclei from the fission products that are formed after all the prompt
neutron and $\gamma$ decays, we still call the precursors ``fission
fragments'' in this paper. Model parameters in BeoH, such as the total
kinetic energy, initial fission fragment mass and spin distributions,
are adjusted to reproduce experimental data of fission product yields,
as well as average prompt and delayed neutron multiplicities, {\it
  e.g.}~\cite{Lovell2022,Lovell2025}. BeoH produces about 10,000
discrete $\gamma$ lines per fission as far as the nuclear structure of
fission fragments is known. Since the BeoH output is normalized to
per-fission, the calculated results are multiplied by the fission
cross section in ENDF/B-VIII.0~\cite{ENDF8} to convert them into the
production cross section. An example of the calculated $\gamma$-ray
production cross section is shown in Fig.~\ref{fig:discgamma}, which
is the 19.3-MeV neutron-induced reaction case. These $\gamma$ lines
are for the discrete transitions only. There are also $\gamma$ rays
produced by the transition between the continuum regions, or from the
continuum to discrete levels.  Since the continuum $\gamma$-ray
spectrum is relatively flat, it forms a base background of the
measured $\gamma$-ray spectrum. On top of that, we see each of the
$\gamma$ rays produced by the (n,$x$n) reactions, which would be
surrounded by many fission $\gamma$ rays.

\begin{figure}
 \resizebox{\columnwidth}{!}{\includegraphics{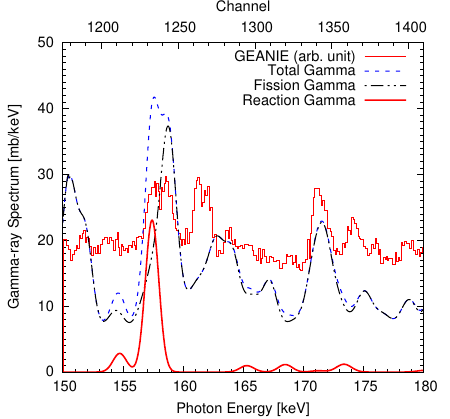}}
 \caption{Calculated $\gamma$-ray spectra from the fission fragments
   (dot-dashed curve), and (n,$x$n$\gamma$) reactions (solid)
   for the neutron incident energy of 11.4~MeV.
   The total spectrum is shown by the dashed curve.
   This figure simulates Figs.~4a in Ref.~\cite{UCRL-ID-140308} before
   background subtraction, which is shown by the histogram.}
 \label{fig:gammaspec_11373}
\end{figure}

The calculated $\gamma$-ray production cross sections are
Gaussian-broadened to convert them into the continuous $\gamma$-ray
spectrum in order to estimate the relative strength of (n,$x$n)
reactions and fission $\gamma$ rays. The broadening width is
empirically taken to be $\Delta = 2\times 10^{-3}\sqrt{E_\gamma}$~MeV,
where $E_\gamma$ is in MeV. This was estimated from Figs.~3 and 4 in
Ref.~\cite{UCRL-ID-140308}. To compare with the GEANIE raw data shown
in Ref.~\cite{UCRL-ID-140308}, the predicted $\gamma$-ray spectrum at
11.373~MeV in the vicinity of 160~keV is shown in
Fig.~\ref{fig:gammaspec_11373}. This simulates Fig.~4a in
Ref.~\cite{UCRL-ID-140308} that shows prominent peaks near the channel
numbers 1240, 1260, and 1340. Although the second peak slightly shifts
higher, our calculation also shows these peaks due to the prompt
$\gamma$ emission from fission fragments. Note that the calculated
fission $\gamma$ spectrum includes the continuum component in order to
compare with the GEANIE data before background subtraction.

\subsection{Data correction}

In addition to the fission $\gamma$ rays, some other (n,$x$n)
reactions may produce $\gamma$ lines very close to the desired line,
albeit rarely.  For example, the 273.4-keV $\gamma$ ray from inelastic
scattering and the 273.3~keV from the (n,3n$\gamma$) reaction may
produce a non-differentiable peak in a measured $\gamma$-ray
spectrum. The contribution of the 273.4-keV line must be subtracted
from the 273.3-keV data to obtain the (n,3n$\gamma$) cross section. In
order to subtract contributions from prompt fission $\gamma$ rays, as
well as other reactions, we calculate the ratio of the $\gamma$-ray
intensity of the net (n,$x$n) reaction to the total intensity
\begin{eqnarray}
  f_c(E_n,E_\gamma)
    &=& \frac{ I_T(E_n,E_\gamma) }{ I_R(E_n,E_\gamma) + I_F(E_n,E_\gamma) } \ , 
  \label{eq:correction} \\
  I_T(E_n,E_\gamma) 
   &=& \int_{E_\gamma - \delta_0}^{E_\gamma + \delta_1} \phi_T(E_n,\epsilon) d\epsilon \ ,\\
  I_R(E_n,E_\gamma)
    &=& \int_{E_\gamma - \delta_0}^{E_\gamma + \delta_1} \phi_R(E_n,\epsilon) d\epsilon \ ,\\
  I_F(E_n,E_\gamma)
    &=& \int_{E_\gamma - \delta_0}^{E_\gamma + \delta_1} \phi_F(E_n,\epsilon) d\epsilon \ ,
\end{eqnarray}
where $E_n$ is the incident neutron energy, $\phi_T(E_n,\epsilon)$ is
the broadened $\gamma$-ray spectrum originated from the target
$\gamma$-line only, $\phi_R(E_n,\epsilon)$ is the spectrum that
includes all the (n,$x$n) reaction $\gamma$ rays, and
$\phi_F(E_n,\epsilon)$ is the spectrum from all fission fragments.
Note that $\phi_F(E_n,\epsilon)$ includes contributions from the
discrete transitions only, since the continuous component is likely
subtracted already at the experimental data analysis as a background.

\begin{figure}
 \resizebox{\columnwidth}{!}{\includegraphics{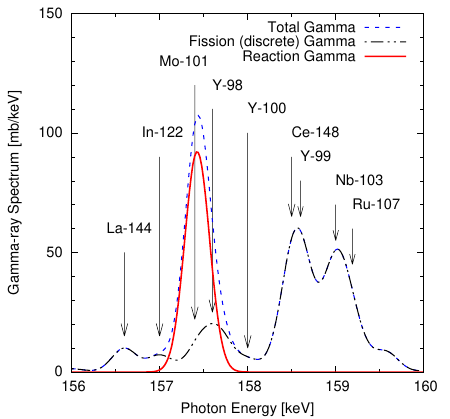}}
 \caption{Calculated $\gamma$-ray spectra from the fission fragments and
   157.4-keV (n,2n$\gamma$) reaction for the neutron incident energy of 11.4~MeV.}
 \label{fig:gammaspec_11373d}
\end{figure}

Unfortunately there was no robust way to determine the integration
range $[E_\gamma-\delta_0$, $E_\gamma+\delta_1]$, because the
background $\gamma$ rays show up randomly around $E_\gamma$. First we
take a narrower energy resolution of $\Delta = 5\times
10^{-4}\sqrt{E_\gamma}$~MeV for the correction factor calculation to
avoid a strong integration range dependence, which eliminates a long
tail from $\gamma$ rays at distant energies. Then within $\delta =
2\Delta \sim 3\Delta$, we determine $\delta_{0,1}$ empirically.  An
example is shown in Fig.~\ref{fig:gammaspec_11373d}, which is
$\phi_R(E_n,\epsilon)$ and $\phi_F(E_n,\epsilon) +
\phi_R(E_n,\epsilon)$ near 157.4~keV for the incident neutron energy
of $E_n=11.4$~MeV. We observed about 50 discrete $\gamma$ lines in the
156--160-keV energy range, and the most prominent lines that add
background to the 157.4-keV line are from $^{122}$In, $^{101}$Mo, and
$^{98}$Y.

The correction factor in Eq.~(\ref{eq:correction}) tends to
over-correct the experimental data, when the fission $\gamma$ spectrum
forms a slope near the target $\gamma$ line, like in
Fig.~\ref{fig:gammaspec_11373d}.  When raw experimental data are
analyzed, it is likely that one may draw a background curve from the
peak of $^{122}$In to somewhere near 157.8~keV, and this
energy-dependent background is subtracted from the total counts. If
this is the case, the fission contribution might be already subtracted
from the data, and we modify the correction factor as
\begin{equation}
  f'_c(E_n,E_\gamma)
   = \frac{ I_T(E_n,E_\gamma) }
          { I_R(E_n,E_\gamma) + I_F(E_n,E_\gamma) - I_B(E_n,E_\gamma)} \ ,
 \label{eq:correction2}
\end{equation}
where the background term $I_B(E_n,E_\gamma)$ is approximated by a trapezoid
\begin{eqnarray}
  I_B(E_n,E_\gamma) 
   &=& \left\{ \phi_X(E_n,E_\gamma - \delta_0) + \phi_X(E_n,E_\gamma + \delta_1)\right\} \nonumber \\
   &\times& \frac{\delta_0 + \delta_1}{2} \ ,
\end{eqnarray}
\begin{equation}
  \phi_X(E_n,\epsilon) = \phi_R(E_n,\epsilon) + \phi_F(E_n,\epsilon) \ .
\end{equation}
In the case of Fig.~\ref{fig:gammaspec_11373d}, the integration range
is from 157.0~keV to 157.8~keV, and Eq.~(\ref{eq:correction}) gives
the correction factor $f_c$ of 0.761, while Eq.~(\ref{eq:correction2})
give a more modest correction of $f'_c = 0.982$. In this case we adopt
$f'_c$.  Depending on how the fission background varies in the
integration range, we adopt either Eq.~(\ref{eq:correction}) or
Eq.~(\ref{eq:correction2}) on a case-by-case basis.  Basically, when
$\phi_R(E_n,\epsilon)$ is on a slope of
$\phi_R(E_n,\epsilon)+\phi_F(E_n,\epsilon)$ like in
Fig.~\ref{fig:gammaspec_11373d}, we employ Eq.~(\ref{eq:correction2}).

Since $\phi_R(E_n,\epsilon)$ is calculated with the CoH$_3$ code, the
correction factor is model-code dependent, and this is a systematic
uncertainty inherent to the method employed here. If we use another
Hauser-Feshbach code, or a different set of model parameters, the
correction factors are no longer the same, nevertheless the difference
should not be so significant as far as the model calculation
reasonably reproduces well known cross sections, such as the total and
fission reactions. However, we would like to emphasize that the
correction factors do not modify the original GEANIE data so as to
coincide with the model calculations. The calculated
$\phi_R(E_n,\epsilon)$ and $\phi_F(E_n,\epsilon)$ are used to estimate
relative contributions of $\gamma$ rays in the spectrum in the
vicinity of the energy of interest, and they are not being optimized
to reduce discrepancies between GEANIE data and Hauser-Feshbach
results.

\section{Results and Discussions}
\label{sec:results}

\subsection{Inelastic scattering $\gamma$ rays}

GEANIE reported four $\gamma$ rays produced by the neutron inelastic
scattering off $^{239}$Pu, whose energies are 154.7, 226.4, 228.2, and
277.6~keV. The discrete levels which produced these $\gamma$ lines are
shown in Table~\ref{tbl:inelgamma}. The branching ratios~\cite{RIPL3}
used in the CoH$_3$ calculations are also shown in this table.

\begin{table}
\caption{$\gamma$ rays reported by the GEANIE experiment
         for neutron inelastic scattering off $^{239}$Pu. }
\label{tbl:inelgamma}
\begin{tabular}{rrcrcr}
\hline
$E_\gamma$ & Initial State &        &  Final State &        & Branching Ratio\\
      keV &           keV & $J^\Pi$ &          keV & $J^\Pi$ & \% \\
\hline
   226.4 &  511.8 & $7/2^+$  & 285.5 & $5/2^+$   & 58.99 \\
   154.7 &  318.5 & $13/2^+$ & 163.8 & $9/2^+$   & 100.0 \\
   228.2 &  285.5 & $5/2^+$  &  57.3 & $5/2^+$   & 43.25 \\
   277.6 &  285.5 & $5/2^+$  &   7.9 & $3/2^+$   & 40.17 \\
\hline
\end{tabular}
\end{table}

\begin{figure}
 \resizebox{\columnwidth}{!}{\includegraphics{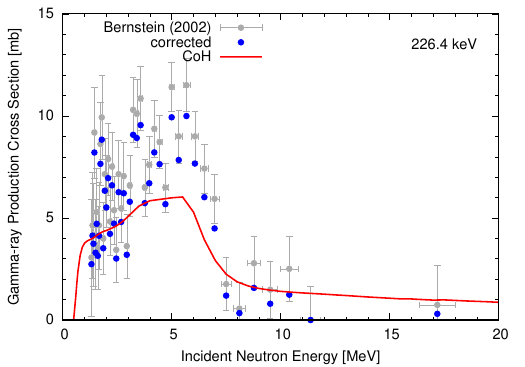}}
 \caption{Calculated 226.4-keV $\gamma$-ray production cross section 
   by the (n,n'$\gamma$) reaction, compared with original and corrected
   GEANIE data.}
 \label{fig:inelastic226keV}
\end{figure}

\begin{figure}
 \resizebox{\columnwidth}{!}{\includegraphics{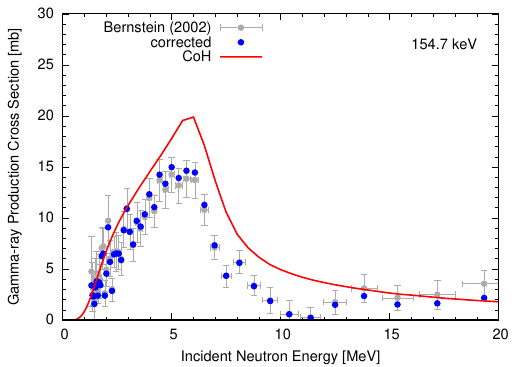}}
 \caption{Calculated 154.7-keV $\gamma$-ray production cross section 
   by the (n,n'$\gamma$) reaction, compared with original and corrected
   GEANIE data.}
 \label{fig:inelastic154keV}
\end{figure}

\begin{figure}
 \resizebox{\columnwidth}{!}{\includegraphics{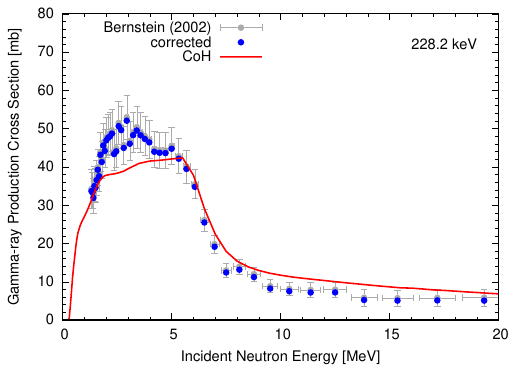}}
 \caption{Calculated 228.2-keV $\gamma$-ray production cross section 
   by the (n,n'$\gamma$) reaction, compared with original and corrected
   GEANIE data.}
 \label{fig:inelastic228keV}
\end{figure}

\begin{figure}
 \resizebox{\columnwidth}{!}{\includegraphics{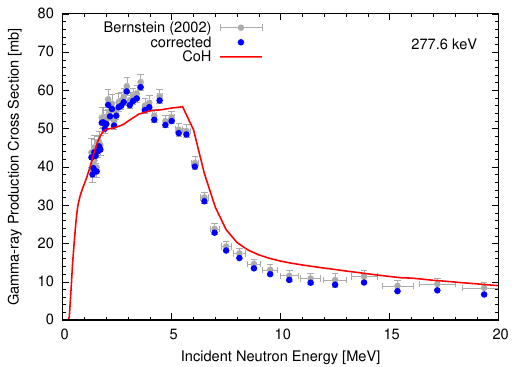}}
 \caption{Calculated 277.6-keV $\gamma$-ray production cross section 
   by the (n,n'$\gamma$) reaction, compared with original and corrected
   GEANIE data.}
 \label{fig:inelastic277keV}
\end{figure}

Figures~\ref{fig:inelastic226keV}--\ref{fig:inelastic277keV} compares
the CoH$_3$ calculated $\gamma$-ray production cross sections with the
original GEANIE data~\cite{UCRL-ID-140308} and the corrected ones.  We
suppressed the error bars of the corrected data for better visibility.
Later we will discuss about the uncertainty of corrected data. In the
inelastic scattering case, the correction factors for the 154.7,
228.2, and 277.6-keV $\gamma$ rays were modest, while the 226.4-keV
$\gamma$ ray was reduced by more than 20\%. One of the reasons for
this large correction is $^{96}$Zr that produces 226.8-keV $\gamma$
line (transition from 3.3092~MeV $6^+$ to 3.0824~MeV $4^+$). The
production cross section of this $\gamma$ ray is about 1--2~mb in the
1--20~MeV energy range.

By adopting these corrected GEANIE data, we can infer the total
inelastic scattering cross section by multiplying the calculated ratio
at the incident neutron energy of $E_n$
\begin{equation}
  r_1(E_n) = \frac{\sigma_{\rm inl}(E_n)}{\sigma_\gamma(E_n)} \ ,
  \label{eq:ratio1}
\end{equation}
where we calculate the total and partial inelastic scattering cross
sections $\sigma_{\rm inl}(E_n)$ and $\sigma_\gamma(E_n)$ with the
CoH$_3$ code.  Because it is very difficult to estimate the
uncertainty of the correction factor, $\delta f_c(E_n,E_\gamma)$, we
assume it to be 10\%. The main contribution from the fission $\gamma$
rays is coming from fission fragments whose yield is relatively
high. Since a typical uncertainty in the evaluated fission product
yield (FPY) data is a few percent when FPY's are in the order of one
percent, the estimated 10\% might be still conservative. In addition,
we add another 5\%, which is empirically estimated for $r(E_n)$.

\begin{figure}
 \resizebox{\columnwidth}{!}{\includegraphics{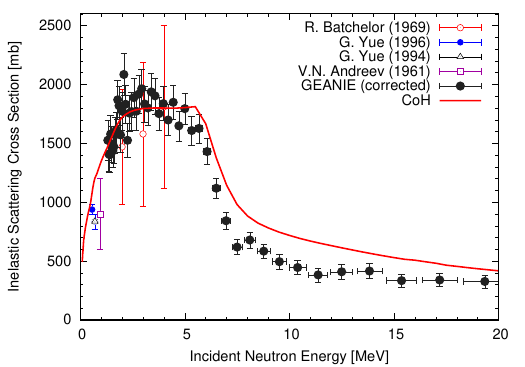}}
 \caption{Total inelastic scattering cross section of $^{239}$Pu inferred
   from the corrected (n,n'$\gamma$) GEANIE data, compared with the CoH$_3$
   calculation. The experimental data by Yue {\it et al.} are corrected by the 
   inelastic scattering to the first excited state.}
 \label{fig:total1n}
\end{figure}

By averaging the corrected four $\gamma$ rays multiplied by $r(E_n)$,
the inferred total inelastic scattering cross section is shown in
Fig.~\ref{fig:total1n}, together with published experimental data;
Batchelor and Wyld~\cite{Batchelor1969}, Yue {\it et
  al.}~\cite{Yue1994, Yue1996}, and
Andreev~\cite{Andreev1963}. Because experimental data by Yue {\it et
  al.} do not include the inelastic scattering to the first excited
state, we calculated these cross sections with CoH$_3$ and added them
to their data. The experimental data by Batchelor and Wyld, and
Andreev are plotted as is, as detailed information about these data is
not available. However, probably these data also do not include the
inelastic scattering to the first level, since the excitation energy
is only 8~keV.

The agreement between the GEANIE/CoH$_3$ inferred data and CoH$_3$
prediction itself seems to be fair up to 5~MeV or so. However they
start deviating above 7~MeV, as all four $\gamma$-ray production cross
sections are already lower than the CoH$_3$ calculations.  It may be
possible to reduce the CoH$_3$ calculation by modifying the model
parameters above 7~MeV. However, this reduction component increases
other reaction channels, which results in an overestimation of the
(n,2n) cross section.

\subsection{(n,2n) Reaction $\gamma$ rays}

There are 8 $\gamma$ rays identified as the (n,2n$\gamma$) reaction,
which are shown in Table~\ref{tbl:2ngamma}.  For the 617.3 and 924-keV
$\gamma$ rays, we see two more transitions whose energies are very
close to each other. Since the difference is less than 1~keV, it is
likely that they overlap. We adopted one of them that has a larger
production cross section as the target $\gamma$ line. For the
617.3-keV data, we assigned this to the transition from the 763.2-keV
level. For the 924-keV data, we took the transition from the 968.2-keV
level.

\begin{table}
\caption{$\gamma$ rays reported by the GEANIE experiment
         for the $^{239}$Pu(n,2n) reaction. The $\gamma$ rays
         in parentheses are treated as background to correct the GEANIE data.}
\label{tbl:2ngamma}
\begin{tabular}{rrcrcr}
\hline
$E_\gamma$ & Initial State &        &  Final State &         & Branching Ratio\\
      keV &           keV & $J^\Pi$ &          keV & $J^\Pi$ & \% \\
\hline
   918.7 &   962.8 &   $ 1^-$ &   44.1 &  $ 2^+$ &  42.55 \\
   924.1 &   968.2 &   $ 2^-$ &   44.1 &  $ 2^+$ &  79.58 \\
 (924.0) &  1069.9 &   $ 3^+$ &  146.0 &  $ 4^+$ &  23.08 \\
   936.6 &  1082.6 &   $ 4^-$ &  146.0 &  $ 4^+$ &  79.69 \\
   962.8 &   962.8 &   $ 1^-$ &    0.0 &  $ 0^+$ &  51.56 \\
   157.4 &   303.4 &   $ 6^+$ &  146.0 &  $ 4^+$ & 100.0 \\
   210.2 &   513.6 &   $ 8^+$ &  303.4 &  $ 6^+$ & 100.0 \\
   459.9 &   763.2 &   $ 5^-$ &  303.4 &  $ 6^+$ &   3.33 \\
   617.3 &   763.2 &   $ 5^-$ &  146.0 &  $ 4^+$ &  96.67 \\
 (617.3) &   661.4 &   $ 3^-$ &   44.1 &  $ 2^+$ &  64.69 \\
\hline
\end{tabular}
\end{table}

\begin{figure}
 \resizebox{\columnwidth}{!}{\includegraphics{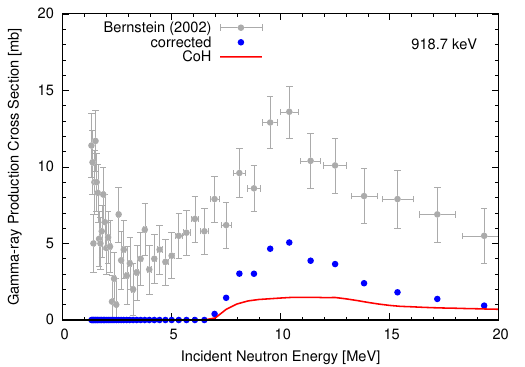}}
 \caption{Calculated 918.7-keV $\gamma$ ray production cross section 
   by the (n,2n$\gamma$) reaction, compared with original and corrected
   GEANIE data.}
 \label{fig:2n919keV}
\end{figure}

\begin{figure}
 \resizebox{\columnwidth}{!}{\includegraphics{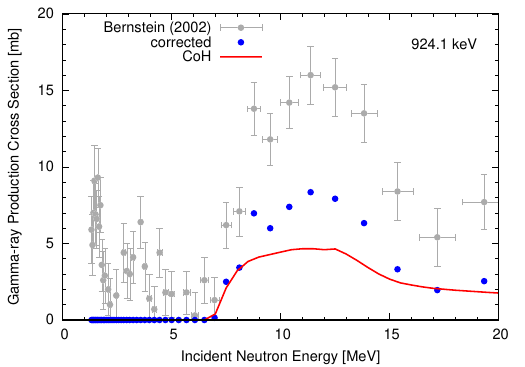}}
 \caption{Calculated 924.1-keV $\gamma$ ray production cross section 
   by the (n,2n$\gamma$) reaction, compared with original and corrected
   GEANIE data.}
 \label{fig:2n924keV}
\end{figure}

\begin{figure}
 \resizebox{\columnwidth}{!}{\includegraphics{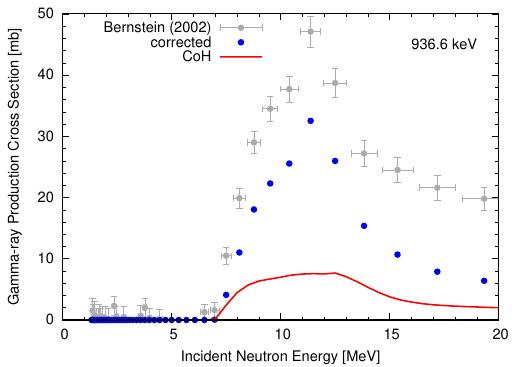}}
 \caption{Calculated 936.6-keV $\gamma$ ray production cross section 
   by the (n,2n$\gamma$) reaction, compared with original and corrected
   GEANIE data.}
 \label{fig:2n937keV}
\end{figure}

\begin{figure}
 \resizebox{\columnwidth}{!}{\includegraphics{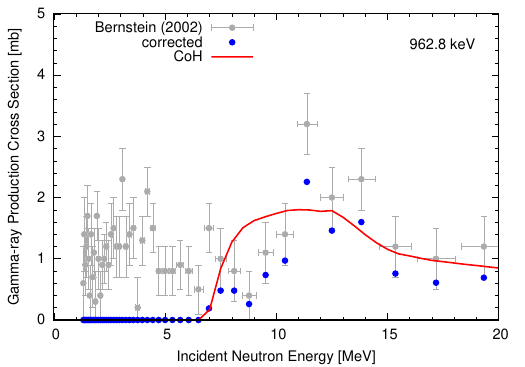}}
 \caption{Calculated 962.8-keV $\gamma$ ray production cross section 
   by the (n,2n$\gamma$) reaction, compared with original and corrected
   GEANIE data.}
 \label{fig:2n963keV}
\end{figure}

\begin{figure}
 \resizebox{\columnwidth}{!}{\includegraphics{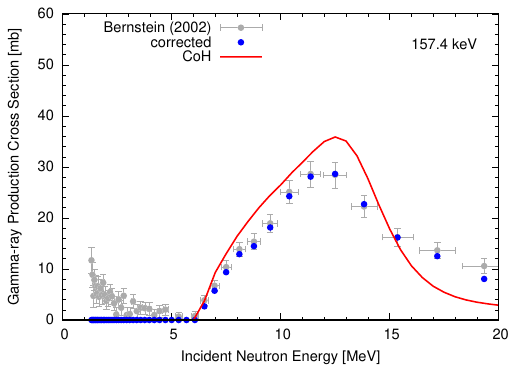}}
 \caption{Calculated 157.4-keV $\gamma$ ray production cross section 
   by the (n,2n$\gamma$) reaction, compared with original and corrected
   GEANIE data.}
 \label{fig:2n157keV}
\end{figure}

\begin{figure}
 \resizebox{\columnwidth}{!}{\includegraphics{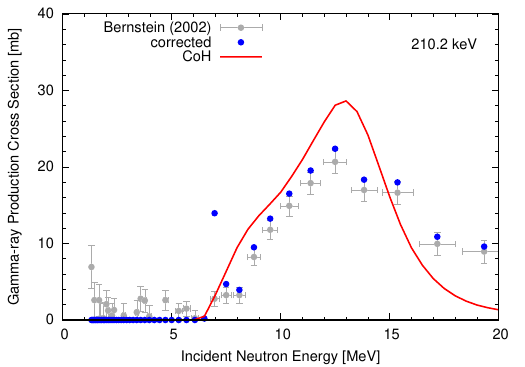}}
 \caption{Calculated 210.2-keV $\gamma$ ray production cross section 
   by the (n,2n$\gamma$) reaction, compared with original and corrected
   GEANIE data.}
 \label{fig:2n210keV}
\end{figure}

\begin{figure}
 \resizebox{\columnwidth}{!}{\includegraphics{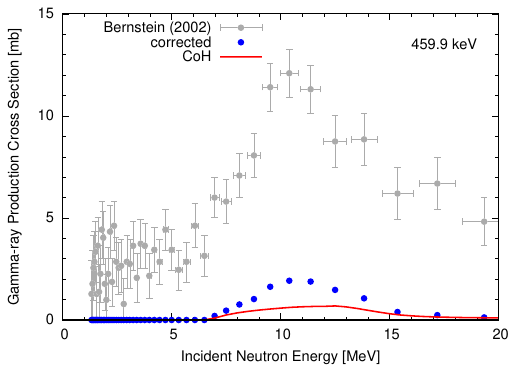}}
 \caption{Calculated 459.9-keV $\gamma$ ray production cross section 
   by the (n,2n$\gamma$) reaction, compared with original and corrected
   GEANIE data.}
 \label{fig:2n460keV}
\end{figure}

\begin{figure}
 \resizebox{\columnwidth}{!}{\includegraphics{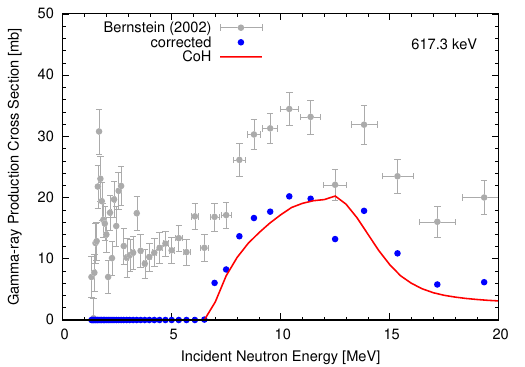}}
 \caption{Calculated 617.3-keV $\gamma$ ray production cross section 
   by the (n,2n$\gamma$) reaction, compared with original and corrected
   GEANIE data.}
 \label{fig:2n617keV}
\end{figure}

Figures~\ref{fig:2n919keV}--\ref{fig:2n617keV} compare the CoH$_3$
calculated $\gamma$-ray production cross sections with the original
GEANIE data and the corrected ones.  Because
Eqs.~(\ref{eq:correction}) and (\ref{eq:correction2}) ensure that
cross section below the threshold energy is zero, unphysical data at
low energies are automatically attributed to the fission $\gamma$ rays
or inelastic scattering, then they are eliminated.

Generally speaking the corrections to GEANIE data are significant, and
the resultant shape of the excitation function becomes much closer to
the statistical model calculation. Because sizable cross sections were
observed below the threshold energy by GEANIE, these $\gamma$ lines by
the fission and/or inelastic scattering likely contaminated the GEANIE
data above the threshold energy too. The corrected data were
significantly reduced by eliminating this contamination.

BeoH produces seven $\gamma$ lines in the energy range from 918 to
920~keV, and three of them produced by $^{139}$Xe and $^{94,96}$Zr
have relatively strong intensities. The original GEANIE 918.7-keV
$\gamma$ ray in Fig.~\ref{fig:2n919keV} is severely contaminated by
these fission $\gamma$ rays. Although the corrected data become closer
to the CoH$_3$ prediction, these data could have very large
uncertainty.


As mentioned before, there are two $\gamma$ lines produced by the
(n,2n$\gamma$) reaction near 924~keV. One of them is considered to
form the background in our treatment. In addition, $^{94}$Sr,
$^{96}$Zr, $^{143}$La, and $^{146}$Ce produce strong $\gamma$ lines in
the vicinity of 924~keV. By subtracting these contributions, the
corrected data agree with the CoH$_3$ result near the threshold
energy, albeit they are still higher above it.


The 936.6-keV $\gamma$ ray in Fig.~\ref{fig:2n937keV} seems relatively
clean as GEANIE gives the correct threshold. However, the corrected
data are still too large compared with the CoH$_3$
prediction. Although BeoH produces more than 10 $\gamma$ lines in the
935--938-keV region, the calculated intensity of the fission $\gamma$
is insufficient to reduce the GEANIE data down to the CoH$_3$
prediction level. The 936.6-keV $\gamma$ ray is a transition from
1082.6~keV $4^-$ to one of the ground state rotational band members of
146.0~keV $4^+$. CoH$_3$ predicts the 1082.6-keV level production
cross section of 8.9~mb at 10~MeV, and its 80\% decays to the $4^+$
level. Roughly speaking, the level production cross sections in the
excitation energy of 0.9--1.2~MeV region are calculated to be 1--10~mb
at 10~MeV.  To obtain the cross section of 25~mb or so for the
936.6-keV $\gamma$ ray, the level production should be 30~mb, which is
presumably impossible in our Hauser-Feshbach calculations, as it would
require a (n,2n) cross section three times larger than the present
result.  There might be some other strong fission $\gamma$-rays that
cannot be seen in the nuclear structure database of BeoH.

The 962.8-keV $\gamma$-ray data are seriously influenced by fission
because the $\gamma$-ray production cross section is in the order of
1~mb.  The corrected data, which are smaller than CoH$_3$, might be
unreliable.

The transitions $8^+ \to 6^+$ and $6^+ \to 4^+$ belong to the ground
state rotational band, and the branching ratio is 100\%.  As the
calculated 210.2-keV $\gamma$ ray ($8^+ \to 6^+$) already exceeds the
original GEANIE data below 15~MeV, the 157.4-keV case is similar.  Our
CoH$_3$ calculation considers a quantum mechanical effect on the spin
transfer in the pre-equilibrium process~\cite{Kawano2006b,
  Dashdorj2007}, which suppresses production of high-spin states such
as $8^+$. If we turn this option off, the calculated cross section
becomes more discrepant with the GEANIE data. This is also seen in the
$^{238}$U(n,n'$\gamma$) data~\cite{Kerveno2021}. It is worth
mentioning that the 210.2-keV $\gamma$ ray is the special case; the
correction factor by Eq.~(\ref{eq:correction2}) became more than
unity, and the corrected data are slightly larger than the original
GEANIE.

The 459.9-keV $\gamma$ ray in Fig.~\ref{fig:2n460keV} is one of the
most striking cases; the reduction in the cross section was
remarkable. The 763.2-keV $5^-$ level production is about 18~mb, and
its 3.33\% produces the 459.9-keV $\gamma$ ray.  To reproduce the
corrected GEANIE data, the branching ratio must be 2--3 times larger
than the reported value of 3.33\%, which might be unlikely. However,
it does not worsen decent agreement of the 617.3-keV $\gamma$ ray in
Fig.~\ref{fig:2n617keV}, which is produced from the same level of
763.2~keV, as the branching ratio is almost 100\%.

\begin{figure}
 \resizebox{\columnwidth}{!}{\includegraphics{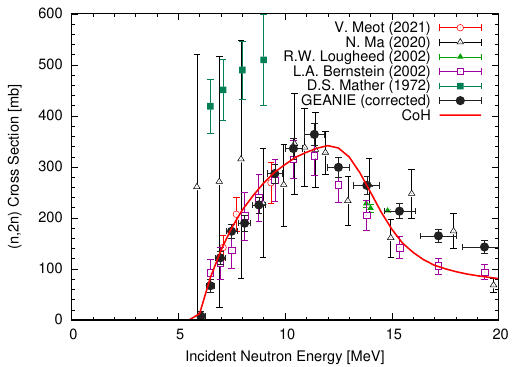}}
 \caption{Total (n,2n) reaction cross section of $^{239}$Pu inferred
   from the corrected (n,2n$\gamma$) GEANIE data, compared with the CoH$_3$
   calculation.}
 \label{fig:total2n}
\end{figure}

By applying the ratio 
\begin{equation}
  r_2(E_n) = \frac{\sigma_{2n}(E_n)}{\sigma_\gamma(E_n)} \ ,
  \label{eq:ratio2}
\end{equation}
we can derive the total (n,2n) cross section. We average all of the
$\gamma$ rays reported by GEANIE, although some of them might be
questionable, to infer the (n,2n) reaction cross section, which is
shown in Fig.~\ref{fig:total2n} with available experimental
data~\cite{Mather1972,Meot2021,Ma2020,Lougheed2002}. The experimental
data of Ma {\it et al.} do not come from a direct measurement but
derived from $^{236}$U($\alpha$,2n) data. The data of Bernstein {\it
  et al.}~\cite{Bernstein2002} are based on the same GEANIE $\gamma$
ray production data but deduced by employing the GNASH code. The
GEANIE/CoH$_3$ inferred data agree well with the data of Meot {\it et
  al.}~\cite{Meot2021} as well as the CoH$_3$ calculation itself below
10~MeV. The data are a little larger than Lougheed {\it et
  al.}~\cite{Lougheed2002} in the 14-MeV energy region, in contrast
the GEANIE/GNASH prediction was systematically lower than this
measurement.

\subsection{(n,3n) Reaction $\gamma$ rays}

\begin{figure}
 \resizebox{\columnwidth}{!}{\includegraphics{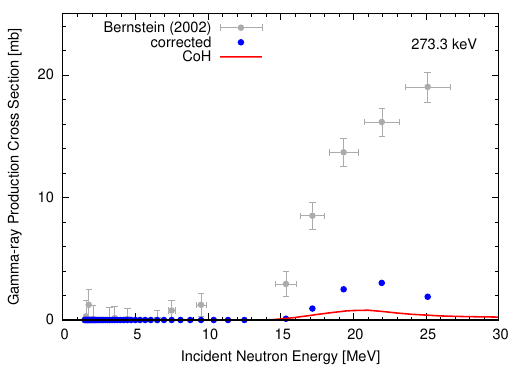}}
 \caption{Calculated 273.3-keV $\gamma$-ray production cross section 
   by the (n,3n$\gamma$) reaction, compared with original and corrected
   GEANIE data.}
 \label{fig:3n273keV}
\end{figure}

The $\gamma$ ray for the (n,3n$\gamma$) reaction is 273.3~keV only, as
shown in Table~\ref{tbl:3ngamma}. The CoH$_3$ calculated results are
compared with the experimental data in Fig.~\ref{fig:3n273keV}.  The
original GEANIE data increase rapidly once the (n,3n) reaction channel
opens. However, the statistical model tells us the production of this
$\gamma$ ray would be less than 1\% of the total (n,3n) reaction at
20~MeV, which means the (n,3n) cross section inferred by the original
GEANIE data exceed unrealistic 1~b. The significant correction made on
the original data implies that these high values are due to other
$\gamma$-ray sources.

\begin{figure}
 \resizebox{\columnwidth}{!}{\includegraphics{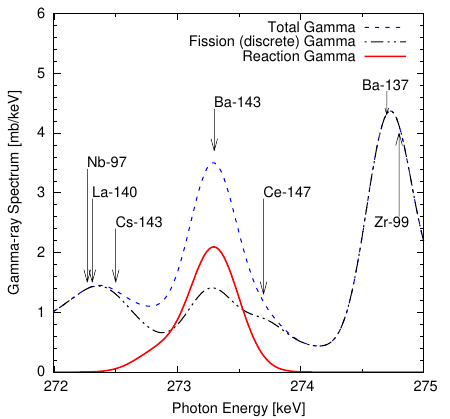}}
 \caption{Calculated $\gamma$-ray spectra from the fission products and
   273.3-keV (n,3n$\gamma$) reaction for the neutron incident energy of 22.0~MeV.
   The reaction component near 273.3~keV includes 273.26-keV inelastic 
   scattering reaction.}
 \label{fig:gammaspec_21960d}
\end{figure}

Figure~\ref{fig:gammaspec_21960d} shows the $\gamma$-ray spectrum near
273~keV for the incident neutron energy of 22~MeV.  The reaction
$\gamma$ component includes the 273.26~keV by the (n,3n) reaction and
273.40~keV by inelastic scattering. The initial state of this
inelastic scattering $\gamma$ is the 779.0-keV $7/2^+$ level, and it
decays to the 505.6~\-keV $5/2^-$ level.  The production cross
sections of these two $\gamma$ rays are of the same order. However,
because the branching ratios from the 779.0-keV level are unknown, we
assumed an equal feeding of 20\% to all the possible final states in
our analysis. If we adopt a larger branching ratio to this decay, the
corrected GEANIE data in Fig.~\ref{fig:3n273keV} become lower.  We
still need accurate branching ratio data in the nuclear structure
database.

The $\gamma$ rays from (n,n') and (n,3n) do not change much as the
incident neutron energy increases from 20 to 25~MeV. The rapid
increase seen in the original GEANIE data might be caused by the
increase in the fission fragment yield of $^{137}$Ba that creates a
long background tail. At the incident energy of 19~MeV, this
$\gamma$-ray production probability is $4.2\times 10^{-4}$ per
fission, while it increases rapidly to $1.0\times 10^{-3}$ at 25~MeV.

\begin{table}
\caption{$\gamma$ rays reported by the GEANIE experiment
         for the $^{239}$Pu(n,3n) reaction.}
\label{tbl:3ngamma}
\begin{tabular}{rrcrcr}
\hline
$E_\gamma$ & Initial State &        &  Final State &         & Branching Ratio\\
      keV &           keV & $J^\Pi$ &          keV & $J^\Pi$ & \% \\
\hline
   273.3 &   321.0 &   $7/2^+$ &  47.7 &  $ 9/2^-$  & 14.45  \\
\hline
\end{tabular}
\end{table}

\begin{figure}
 \resizebox{\columnwidth}{!}{\includegraphics{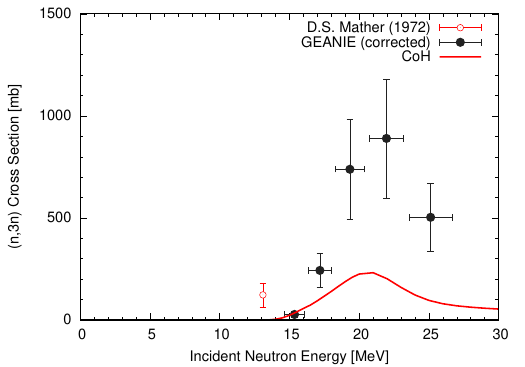}}
 \caption{Total (n,3n) reaction cross section of $^{239}$Pu inferred
   from the corrected (n,3n$\gamma$) GEANIE data, compared with the CoH$_3$
   calculation.}
 \label{fig:total3n}
\end{figure}

The GEANIE/CoH$_3$ inferred (n,3n) cross section is shown in
Fig.~\ref{fig:total3n}. We assumed an extra uncertainty of 30\% for
the branching ratio data. The optical model gives the total reaction
cross section of 2900~mb at 20~MeV, and 2300~mb goes to the fission
channel. Based on the fact that the rest of 600~mb must be shared by
the (n,n'), (n,2n), and (n,3n) channels, the GEANIE/CoH$_3$ cross
section could be too large. As seen in Fig.~\ref{fig:3n273keV}, an
extreme correction was made on the 173.4-keV $\gamma$ ray, our
uncertainty assessment for the total (n,3n) might be still
underestimated.

\section{Conclusion}
\label{sec:conclusion}

The historic GEANIE experiment of $^{239}$Pu(n,$x$n$\gamma$) reaction
was an important milestone for a new technique to derive nuclear
reaction cross sections by combining experimental partial data with
theoretical calculations.  Since then, some of the underlying
technology have matured enough to warrant revisiting this challenging
data processing, which motivated the re-analysis of the GEANIE
data. By aggregating all the $\gamma$ rays produced by fission, as
well as other known transitions in the (n,$x$n) reactions, we
estimated the partial $^{239}$Pu(n,n'), (n,2n), and (n,3n)
$\gamma$-ray production cross sections, which could be less influenced
by the background components. The prompt fission $\gamma$ rays were
calculated with the Hauser-Feshbach fission fragment decay (HF$^3$D)
model code BeoH, which produced more than 10,000 discrete $\gamma$
lines. The (n,$x$n$\gamma$) reactions were calculated with the
statistical Hauser-Feshbach code, CoH$_3$.  When these contributions
were subtracted from the original GEANIE data, agreement between
GEANIE and the statistical Hauser-Feshbach model predictions were
significantly improved.  Based on the corrected partial $\gamma$-ray
production cross sections, we inferred estimates of the total (n,n'),
(n,2n), and (n,3n) cross sections.  The derived (n,n') and (n,2n)
cross sections agree with the published data, while (n,3n) may still
have an unknown strong background that is needed to reconcile the data
with the theoretical prediction. Because using prompt $\gamma$-ray
measurements is a promising technique to estimate reaction cross
sections that are not so easy to measure directly, we conclude that
our data processing procedure will be useful to perform new
experiments for actinides where fission $\gamma$ rays could represent
a significant background.

\begin{acknowledgements}
TK would like to thank M. Kerveno of U. Strasbourg for valuable
discussions.  TK and AL were partially supported by the Office of
Defense Nuclear Nonproliferation Research \& Development (DNN R\&D),
National Nuclear Security Administration, U.S. Department of
Energy. This work was carried out under the auspices of the National
Nuclear Security Administration of the U.S. Department of Energy at
Los Alamos National Laboratory under Contract No. 89233218CNA000001.
\end{acknowledgements}

\appendix
\section{Inferred $^{239}$Pu(n,$x$n) cross sections}
\label{sec:appendix}

\begin{table}
\caption{GEANIE/CoH$_3$ inferred (n,n') cross section}
\begin{tabular}{rrr}
\hline
$E_n$  & Cross Section & Uncertainty \\ 
 MeV   & mb            & \% \\
\hline
   1.321 &      1526 &   11.1\\
   1.365 &      1411 &   10.7\\
   1.409 &      1403 &   10.6\\
   1.455 &      1579 &   9.96\\
   1.504 &      1496 &   10.2\\
   1.556 &      1436 &   9.80\\
   1.610 &      1471 &   9.83\\
   1.667 &      1572 &   9.70\\
   1.728 &      1617 &   9.26\\
   1.792 &      1871 &   8.79\\
   1.858 &      1820 &   8.95\\
   1.929 &      1572 &   9.45\\
   2.004 &      1773 &   8.94\\
   2.084 &      2085 &   8.56\\
   2.169 &      1832 &   8.95\\
   2.259 &      1526 &   9.59\\
   2.354 &      1757 &   8.67\\
   2.455 &      1770 &   9.04\\
   2.564 &      1889 &   8.65\\
   2.680 &      1773 &   8.97\\
   2.802 &      1919 &   8.31\\
   2.938 &      1964 &   8.57\\
   3.078 &      1835 &   8.30\\
   3.231 &      1801 &   8.30\\
   3.396 &      1937 &   8.14\\
   3.574 &      1904 &   8.07\\
   3.766 &      1753 &   8.09\\
   3.973 &      1837 &   7.87\\
   4.200 &      1698 &   7.74\\
   4.445 &      1849 &   7.60\\
   4.713 &      1649 &   7.73\\
   5.001 &      1795 &   7.31\\
   5.325 &      1609 &   7.58\\
   5.678 &      1627 &   7.52\\
   6.064 &      1431 &   7.62\\
   6.500 &      1119 &   7.92\\
   6.969 &     842.8 &   8.58\\
   7.498 &     619.1 &   10.2\\
   8.105 &     679.2 &   9.95\\
   8.773 &     584.6 &   10.6\\
   9.520 &     495.8 &   12.5\\
  10.392 &     446.1 &   13.6\\
  11.373 &     383.1 &   15.7\\
  12.499 &     409.2 &   16.4\\
  13.824 &     415.5 &   15.6\\
  15.359 &     334.4 &   16.3\\
  17.183 &     340.1 &   16.8\\
  19.336 &     327.7 &   16.8\\
\hline
\end{tabular}
\end{table}

\begin{table}
\caption{GEANIE/CoH$_3$ inferred (n,2n) cross section}
\begin{tabular}{rrr}
\hline
$E_n$  & Cross Section & Uncertainty \\ 
 MeV   & mb            & \% \\
\hline
   6.064 &   7.642 &    121\\
   6.500 &   67.07 &   19.5\\
   6.969 &   121.9 &   11.0\\
   7.498 &   173.9 &   8.32\\
   8.105 &   189.7 &   8.06\\
   8.773 &   225.5 &   7.13\\
   9.520 &   287.0 &   6.53\\
  10.392 &   335.9 &   6.19\\
  11.373 &   363.8 &   6.01\\
  12.499 &   299.4 &   6.49\\
  13.824 &   263.7 &   6.54\\
  15.359 &   213.4 &   6.99\\
  17.183 &   165.0 &   8.07\\
  19.336 &   143.2 &   8.46\\
\hline
\end{tabular}
\end{table}

\begin{table}
\caption{GEANIE/CoH$_3$ inferred (n,3n) cross section}
\begin{tabular}{rrr}
\hline
$E_n$  & Cross Section & Uncertainty \\ 
 MeV   & mb            & \% \\
\hline
  15.359 &    27.16 &  46.9\\
  17.183 &    242.8 &  34.5\\
  19.336 &    738.7 &  33.1\\
  21.960 &    889.8 &  32.8\\
  25.121 &    503.5 &  32.7\\
\hline
\end{tabular}
\end{table}

\bibliography{ref}

\end{document}